\documentstyle[12pt,fleqn,epsf]{article}

\newcommand{\bi}{\begin{itemize}}
\newcommand{\ei}{\end{itemize}}

\newcommand {\dzero}	  {D\O}

\newcommand {\met}	  {\mbox{${\rlap{\kern0.25em/}E_{\bot}}$}}

\newcommand {\bbar}	  {\mbox{${\rm\bar{b}}$}}

\newcommand {\ppbar}	  {\mbox{${\rm p\bar{p}}$}}
\newcommand {\qqbar}	  {\mbox{${\rm q\bar{q}}$}}
\newcommand {\bbbar}	  {\mbox{${\rm b\bar{b}}$}}
\newcommand {\ttbar}	  {\mbox{${\rm t\bar{t}}$}}

\newcommand {\rargap}	  {\mbox{ $\rightarrow$ }}

\newcommand {\vtb}	  {\mbox{${V_{\rm tb}}$}}

\newcommand {\tev}	  {\mbox{${\mbox{\,\rm TeV}}$}}
\newcommand {\gev}	  {\mbox{${\mbox{\,\rm GeV}}$}}

\newcommand {\sqs}        {\mbox{${\sqrt{s}}$}}
\newcommand {\pb}         {\mbox{${\,\rm pb}$}}
\newcommand {\ipb}        {\mbox{${\,{\rm pb}^{-1}}$}}
\newcommand {{\wtb}}        {\mbox{${\rm Wtb}$}}
\newcommand {\jbbbar}     {\mbox{${\rm jb\bar{b}}$}}
\newcommand {\jjbbbar}    {\mbox{${\rm jjb\bar{b}}$}}
\newcommand {{\wjj}}        {\mbox{${\rm Wjj}$}}
\newcommand {{\wbbbar}}     {\mbox{${\rm W^\pm b\bar{b}}$}}

\begin{document}
\vspace*{0.8cm}
\begin{center}

{\Large \bf SINGLE TOP QUARK AND LIGHT HIGGS BOSON AT TEVATRON}

\vspace{1cm}

{\underline{A.~Belyaev}, E.~Boos, and L.~Dudko}

{\it Moscow State University, Institute of Nuclear Physics, RU-119899
Moscow, Russia

E-mail: belyaev@sgi.npi.msu.su, boos@theory.npi.msu.su,
dudko@sgi.npi.msu.su}

\end{center}

\vspace{6cm}

\begin{abstract} Search for single top quark and Higgs boson production  is among the main
goals for the upgraded Tevatron. 
We study the feasibility of single top quark and Higgs boson search at
Tevatron both together since these processes have very  similar final state signatures
and so backgrounds.
Calculation of the  cross sections and 
a detailed analysis of
various  kinematic  distributions for signal and  for main backgrounds has been performed. 
  As a result of
this 
study an opimized set
of cuts has been worked out for background suppression and for extraction of the signal. This
study gives proof to the possibility of testing the nature of the {\wtb} coupling and of a
direct measurement of the {\vtb} CKM parameter. The  \bbbar -- pair effective mass
distribution study allows to  extract the signal of Higgs with mass up to about  $100\gev$  at
$\sqs = 2\tev$ and up to about $120\gev$  at $\sqs = 4\tev$ for integrated luminosity $L=
1000\ipb$.

\end{abstract}

\vspace*{-0.9cm}
\section{Single top quark}
\vspace*{-0.3cm}
The top quark production in the QCD {\ttbar} pair channel has been
established by  {\dzero} and CDF collaborations at TEVATRON in
1995\cite{tt}.  The cross section of  electroweak process of single top
quark production  is comparable with {\ttbar} production
(see refs. 8-21 in \cite{our-st}). Single top production diagrams 
directly include {\wtb} coupling, in contrast to {\ttbar}
channel. Therefore, single top production provides a unique opportunity
to study the {\wtb} structure and to measure {\vtb}.
Top quark is heavy and  single top quark study could also be promising in
looking for deviations from the Standard Model (SM).  Despite the cross
section of single top quark is of order of $2\pb$  the task of the
background reduction is much more serious problem then that in  the
{\ttbar} case.  That is why we concentrated here on the possibility to
extract the signal of single top quark from various backgrounds.  The
top quark decays into $\rm W^+$~boson and $\rm b$~quark. We consider
here only subsequent leptonic decays of the $\rm W$ to positron (muon)
and neutrino, as this signal must be easier to find experimentally than
channels with hadronic decays of $\rm W$-boson.
\twocolumn
\vspace*{-1.3cm}
\subsection{Signal}
\begin{figure}[htb]
    \vspace*{-1.4cm}
    \hspace*{-0.3cm}
  \epsfxsize=9cm  
    \epsffile{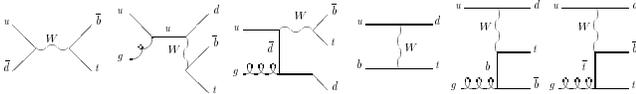}
    \vspace*{-9.9cm}
    \caption{\label{diag-st} Diagrams for single top production}   
\vspace*{-0.3cm}
\end{figure}

\vspace*{-0.2cm}
We concentrated on the following processes of single top
quark production:
1.~${\ppbar}{\rargap}\rm tq{\bbar}+\rm X$ (W-gluon fusion), 
2.~${\ppbar}{\rargap}\rm t{\bbar}+\rm X$, 
3.~${\ppbar}{\rargap}\rm tq+X$
where $\rm q$ is a light quark and $\rm X$ stands for remnants of  the
proton and antiproton. Feynman diagrams for the above
processes are shown
in Fig.~\ref{diag-st}. We refer to the paper \cite{our-st} for details.
All analytical and almost all numerical calculations have been done  by
means of CompHEP package~\cite{comp}. FORTRAN codes of matrix elements
generated by CompHEP have been used in MC generators implemented as
external processes of PYTHIA program. For t-quark production we chose
$Q^2$ scale equal to top mass squared.  For a $180\gev$ top quark   the
total single top plus anti-top cross section is
$1.76_{-0.18}^{+0.26}\pb$ ($\sqs=1.8\tev$). The upper and lower bounds
come from the difference between calculations using two parton
distributions with the errors caused by the choice of QCD scale. The
final state signature of signal for leptonic decay of W is $\rm
e^{+}(\mu^{+})+ \met +2(3)\,jets$ where  one of  the jets is b quark
from top.
\vspace*{-0.4cm}
\subsection{Main backgrounds}
\vspace*{-0.2cm}
The main backgrounds leading to the same final state signature as in
case of single top productions are:  $\ppbar\rargap\rm W+2(3)\,jets$
(gluonic,  $\alpha\cdot\alpha_{\rm s}$ order and electroweak $\alpha^2$
order) and   QCD (${\jbbbar}+{\jjbbbar}$)  background when  one of the
jets imitates electron.

All numbers  both for signal and background are presented below  with
the initial general cuts on jets: $\Delta R_{\rm jj(ej)} > 0.5$,
$p_\bot(\rm jet) > 10\gev$  ($|\Delta R_{\rm
jj}|=\sqrt{\Delta\phi^{2}_{\rm jj}+ \Delta\eta^{2}_{\rm jj}}$).  For
$\rm Wjj$ background $Q^2=M_{\rm W}^2$ has been taken. For calculation
of {\jbbbar} and {\jjbbbar} processes we choose invariant \bbbar mass
for $Q^2$ scale.

Total cross section of  $\rm W+2jets$ ({\wjj}) ``gluonic'' background
($1000\pb$) is about 2  orders of magnitude higher  than that of the
signal.  The specific feature of single top production is highly
energetic $\rm b$ quark in the final state and  one additional $\rm b$
quark for W-gluon fusion process. The $\rm b$ quark content of the
{\wjj} processes is small (less then 1\%) and efficient b-tagging gives
us  chance to extract the signal from {\wjj} overwhelming background.
We assume 50\%  b-tagging efficiency hereafter. For  initial cuts
mentioned above  the total cross section  for $\rm W^+\bbbar$  process
(gluon splitting) is $7.2\pb$. Contribution from {\wjj} background due
to $\rm b$ quark misidentification  should also be taken into account.
In our study  we chose 1\% misidentification probability.

\begin{figure}[htb]
  \vspace*{-1.0cm}
  \hspace*{-0.5cm}
  \epsfxsize=10cm  
  \epsffile{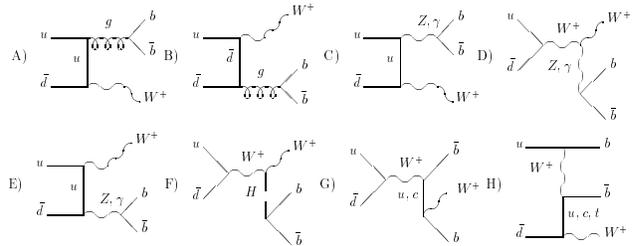}
  \vspace*{-9.2cm}
  \caption{\label{back1} Diagrams for $Wb\bar{b}$ background}
  \vspace*{-0.1cm}
  \end{figure}

Complete set of Feynman diagrams for {\wbbbar}  background is  shown
in  Fig.~\ref{back1}. 
The main contribution to the background  comes from processes A and B
with $\rm g\to \bbbar$ splitting. Diagrams with virtual photon and Z
boson (C, D and E) give small contribution to the total cross section
(for virtual photon less than 1\%) due to presence of additional
electromagnetic constant, while contribution from $\rm WZ$ can be
removed  by imposing cut on the invariant {\bbbar} mass. Process  F
with Higgs production contributes also less than 1\% ($m_{\rm
H}>100\gev$) of the total  {\wbbbar} background. Diagrams G and H are
negligible  due to small value of the CKM elements. 

Another important irreducible background comes from  multi-jet QCD
processes when jet is misidentified as an electron. Though the
probability of that is small (of order of 0.01\%) the cross section for
such processes is large and  gives essential contribution to the
background for single top production. We have calculated total cross
sections and made MC simulations  for {\jbbbar} and {\jjbbbar}
processes  which are relevant for the considered signature if one light
jet imitates an electron. Table~\ref{tab-jbb} summarizes all
subprocesses giving {\jbbbar} and {\jjbbbar} final state signatures and
their cross sections. Total cross sections for {\jbbbar} and {\jjbbbar}
are 204105 and $66733\pb$, respectively. 
\vspace*{-0.3cm}
\begin{table}[htb]
\begin{tabular}{ l r   l r  }
 process & CS(pb)&  process & CS(pb)\\
$g	g	\to g		 b \bar{b}$&  120180 &      
$g 	q \to    q    	 	 b \bar{b}$&  82290    \\
$q\bar{q} \to g		         b \bar{b}$&  1635    & 
$gg  \to gg 			 	 b \bar{b}$& 36212  \\ 
$q\bar{q}  \to  q \bar{q} 		 b \bar{b}$&  2269  & 
$q\bar{q}  \to  gg  			 b \bar{b}$&   288   \\
$gq\to gq				 b \bar{b}$& 26166  & 
$gg  \to q \bar{q} 			 b \bar{b}$&  1798  
\end{tabular}
\vspace*{-0.3cm}
\caption{\label{tab-jbb}  $jb\bar{b}$ and
$jjb\bar{b}$ cross sections.
processes}
\vspace*{-0.2cm}
\end{table}
\vspace*{-0.4cm}
\subsection{Signal and background study}
\vspace*{-0.1cm}
Even after  b-tagging procedure the signal is more than one order lower
than the background. This fact requires special kinematic analysis in
order to reduce the background and leave  the signal intact as much
as possible.

\begin{figure}[htb]
   \begin{center}
    \vskip -0.5cm\hspace*{-0.3cm}
    \epsfxsize=4.5cm\epsffile{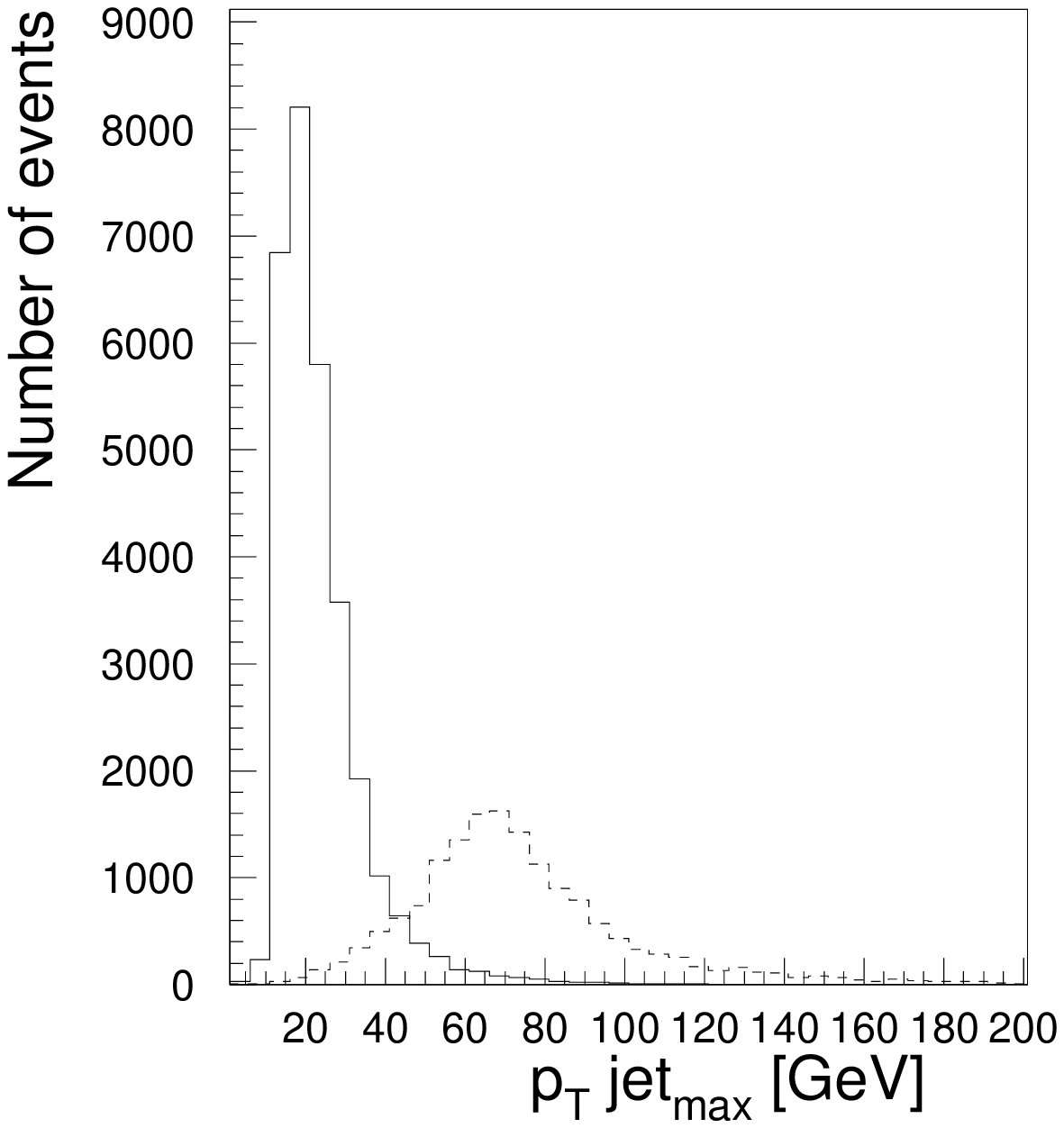}\epsfxsize=4.5cm\epsffile{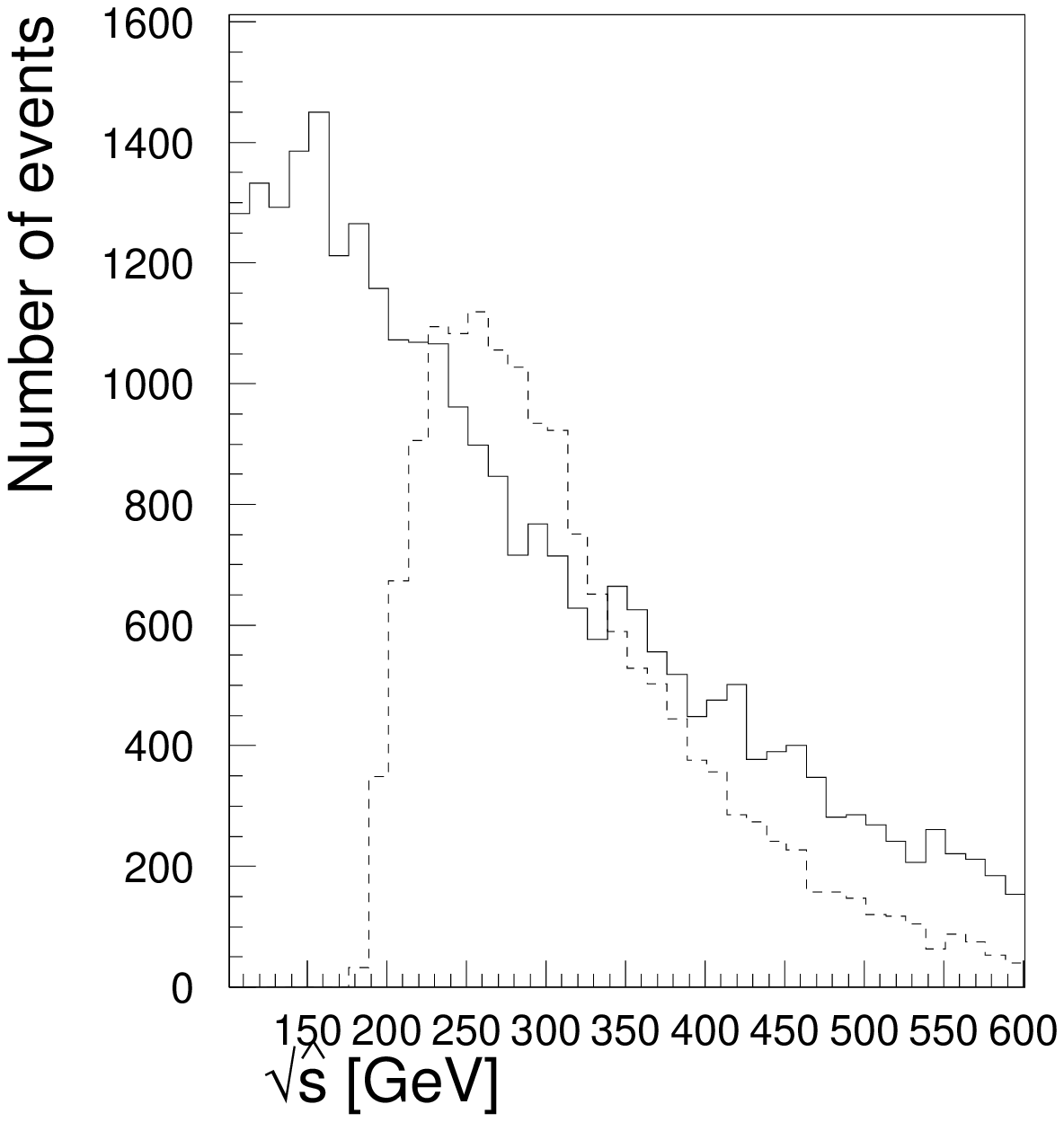}
    \put(-210,100){\tiny --------- background}
    \put(-210,110){\tiny - - - - - 200$\cdot$signal}
    \put(-215,120)  {a)}
    \put(-85, 120)  {b)}
    \put(-215,0)  {c)}
    \put(-85, 0)  {d)}
    \vskip -0.9cm\hspace*{-0.3cm}
    \epsfxsize=4.5cm\epsffile{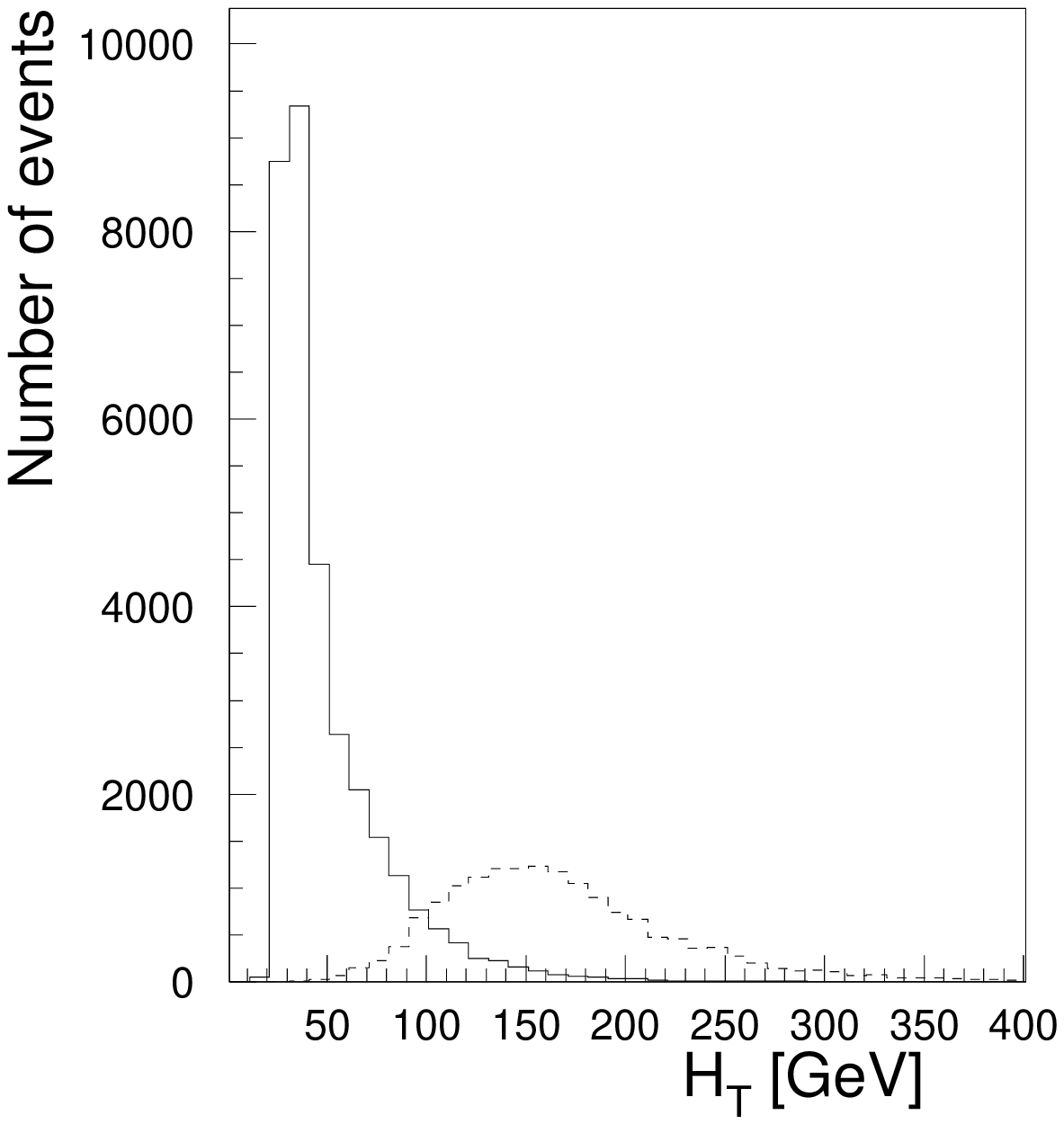}\epsfxsize=4.5cm\epsffile{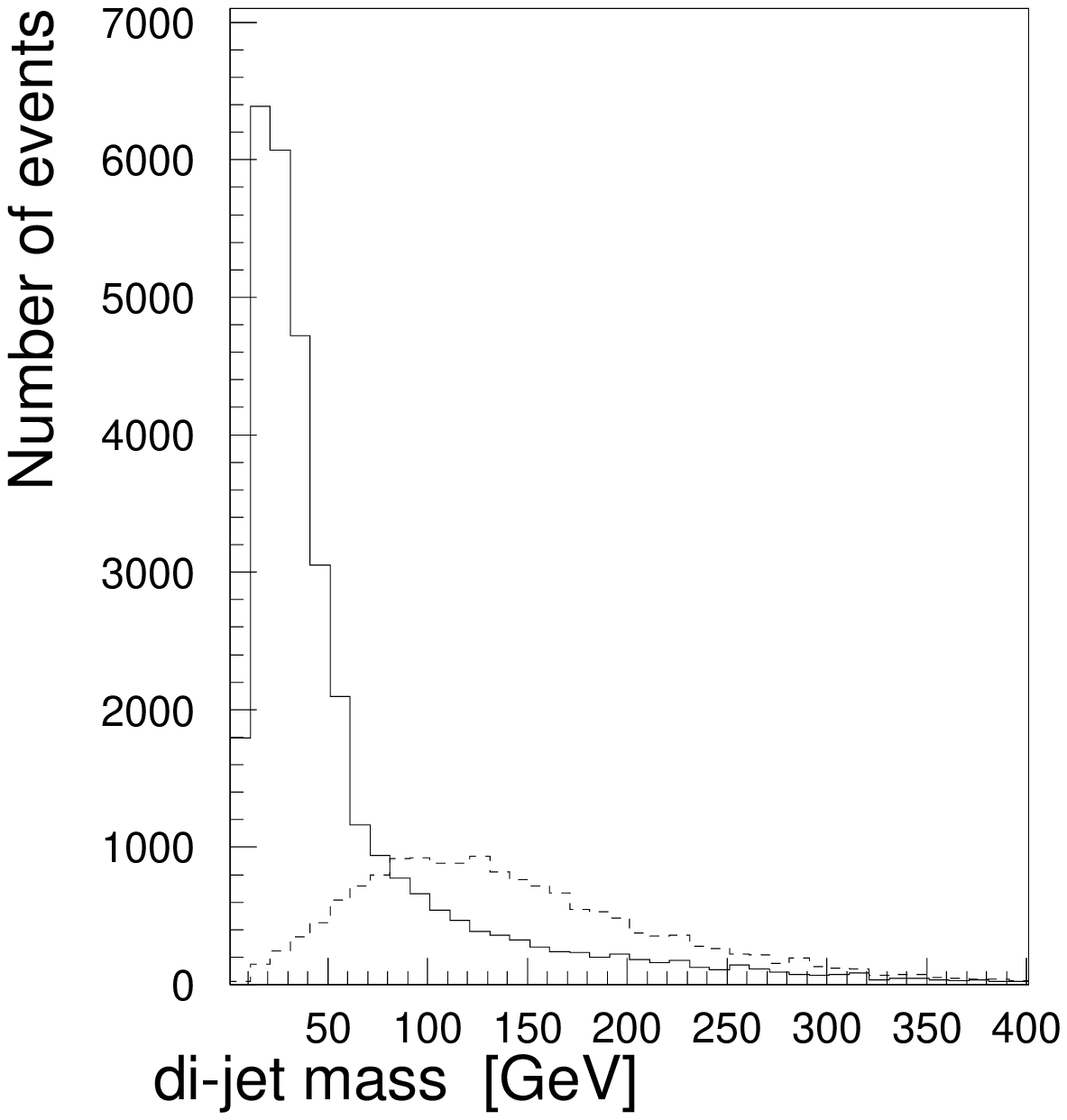}
 \end{center}
  \vspace*{-1.3cm}
  \caption{\label{st-fig1} Distributions for signal and background
  for some most spectacular  variables} 
  \vspace*{-0.3cm}
\end{figure}

The main difference between  kinematic distributions for  background 
and  signal is  that jets from {\wjj}, {\jbbbar} and {\jjbbbar} 
processes are  much softer and less central than  those for the signal
with one very hard jet coming from top and another softer  jet,
accompanying top quark.  Among the kinematic variables for separation
of the signal and background the most effective are: $p_\bot$ of
leading jet (Fig.~\ref{st-fig1}a)  invariant mass of the system
$\sqrt{\hat{s}}$ (Fig.~\ref{st-fig1}b),  which is always greater than
$m_{\rm top}$ for the signal;   $p_\bot({\rm W})$;  scalar transverse
energy $H_\bot$ (Fig.~\ref{st-fig1}c)   ($H_\bot = |E_\bot({\rm jet1})|
+ |E_\bot({\rm jet2})| + |E_\bot({\rm lepton})|$), di-jet mass
(Fig.~\ref{st-fig1}d).
{\small
\begin{table}[htb]
\hspace*{-0.6cm}
\begin{tabular}{ l r r r r   }
\hline
accompanying cuts 		& Top &	$Wb\bar{b}$ &	$Wjj$ &	$j(j)b\bar{b}$
\\
\hline
1. $\Delta R_{jj} >0.5$ ,   &&&&\\
  $P_{Tj} > $ 10 GeV        &    78&             357&	   493&   	25246\\
2. $\met,{p_t}_e>20$~GeV  &    56&		 266&	  370 &    	  289\\
3. $p_{Tj1} > 45$ GeV  &	 49&		  64&	    85&            38\\
4. $\sqrt{\hat{s}} > $180 GeV &  49&		  57&	    43&       	   37\\	
   
5. $p_T W > $30 GeV     &        45&		  50&	    39&       	   37\\ 
 		      
6. $jj_m$$ >$ 25 GeV       &     45&	 	  48&	    38&       	   37\\ 
 		      
7. $H_T > $100 GeV &             44&	  	  46&	    36&       	   36\\ 
 	 	      
\end{tabular}
\vspace*{-0.3cm}
\caption{\label{st-tab} Signal and background events;
\ \ \ \  \ \ \ \ \ $\epsilon_b=50\%$, $m_{top}$=180~GeV, $L$=1000~pb$^{-1}$ }
\vspace*{-0.0cm}
\end{table}}
Based on  different behaviors of signal and background kinematics  the
optimal set of cuts for the background suppression has been worked out.
The effect  of consequent application of these cuts is presented in
Table~\ref{st-tab}.
As a result the signal-to-background ratio becomes equal to 1/3 and the
signal exceeds the background by 3 standard deviations.
%
\subsection{{\wtb} coupling and \vtb}
%
Since the top quark is rather heavy one could expect that new physics
might be revealed at the scale of its mass (see refs. 65-71 in
\cite{our-st}).  We examine the effects  of a deviation in the {\wtb}
coupling from the standard model structure, and study sensitivity of 
measurement of the CKM matrix element {\vtb}. Experimental studies of this
type are among the main goals of the single top physics. 

As an example of a deviation from the standard model {\wtb} coupling, we
introduce additional contribution from the nonstandard 
{\small $(V+A)$} structure
with arbitrary parameter $A_r$ by:\\
$$\Gamma = \frac{eV_{\rm tb}}{2\sqrt{2}\sin\theta_{\rm W}}
                  \left[\gamma_{\mu}\left(1-\gamma_5\right)
                  + A_r\gamma_{\mu}\left(1+\gamma_5\right)\right]$$
The production rate varies almost quadratically with $A_r$, and is
nearly symmetric around $A_r=0$. The cross section rises from $2.44\pb$
when $A_r=0$ to $4.68\pb$ when $A_r=-1$ and to $4.73\pb$ when $A_r=+1$.
We have calculated the region in the ({\vtb},$A_r$) plane where future
single top measurements are expected to be sensitive.  The error of
{\vtb} measurement will be half of the error of the single top cross
section, since the cross sections for all single top processes are
proportional to $|{\vtb}|^2$. This results in an error for {\vtb} and
$A_r$ around 12-19\% ($L=2\,{\rm fb}^{-1}$)\cite{our-st}.
\vspace*{-0.3cm}
\section{Higgs boson}
\vspace*{-0.2cm}
Luminosity upgrade of  TEVATRON and  installation of the  efficient 
b-tagging system opens also  opportunities for the Higgs boson search.
This task is crucial since the Higgs boson is the last  particle (in
the frame of SM)  which has not been discovered yet.

The most promising will be the Higgs boson search in association with
the electroweak W and Z bo\-so\-ns~\cite{our-higgs,will}:\\
\begin{figure}[htb]
\hspace*{0.5cm}$p \bar p \rightarrow W^\pm H + X$\\
\hspace*{0.5cm}$p \bar p \rightarrow Z H + X    $ 
  \vspace*{-6.5cm}
  \begin{center}
  \hspace*{-6.0cm}
    \epsfxsize=20cm
    \epsffile{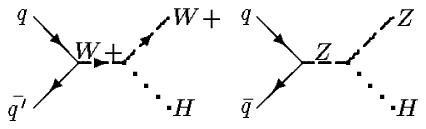}
    \vspace*{-20.7cm}
  \end{center}
\end{figure}%
Here  we concentrated on detailed study  of Higgs signal  (for $M_{\rm
H} \le 140\gev$)   as well as of the background processes including
analysis of the 4-fermion final states distributions. This study shows
possibility of  extraction of the Higgs signal from the background.
\vspace*{-0.4cm}
\subsection{Signal}
\vspace*{-0.2cm}
For dominating decay mode of the Higgs  in the mass interval, $M_{H}
\leq 140\gev$, Higgs production reactions  lead to the  following
signatures: $2\,{\rm b-jets} + \ell + \met + {\rm X}$, $2\,{\rm b-jets}
+ \met + {\rm X}$, $2\,{\rm b-jets} + \ell^{+}\ell^{-} + {\rm X}$,
$2\,{\rm b-jets} + 2\,{\rm jets} + {\rm X}$, where $\ell$ is charged
lepton ($e, \mu, \tau$), $\met$ - missing transverse energy. The latter
reaction is more complicated for calculation and for  the experimental
search, so we do not consider it. The complete set of tree level
Standard Model diagrams contributing  to the final states mentioned
above has been calculated. It includes the following numbers of $2\to
4$ diagrams:\\
 18  - for  $\qqbar \to \bbbar \ell^{\pm}\nu$;
 33  - for   $\qqbar \to \bbbar \ell^{+}\ell^{-}$;\\
 16  - for  ${\rm gg} \to \bbbar \ell^{+}\ell^{-}$; 
 15  - for  $\qqbar \to \bbbar\nu\bar{\nu}$;\\
and 8 diagrams for  ${\rm gg} \to \bbbar\nu\bar{\nu}$.

At $\sqs=2\tev$ the Higgs production cross section is about 0.03-
$0.3\pb$ depending on the Higgs mass and therefore one can expect
30-300 signal events for integrated luminosity $L = 1000\ipb$. However 
the total reaction rates are about two orders of magnitude larger that
the Higgs signal. That is why one has to make a detailed analysis of 
different 4-fermion final state distributions to find out whether it is
possible to find a set of cuts to suppress the huge background strongly
enough. For the cross section calculation and event simulation the same
procedure has been done as for the single top quark study, including
the effect of initial and final state radiation.
\vspace*{-0.4cm}
\subsection{Background study}
\vspace*{-0.2cm}
From the complete set of diagrams for background the dominant
contribution  comes from several subprocesses:
1) QCD $2\to 3$ type main background subprocesses:
$\qqbar^{'}\to {\wbbbar}$,
$\qqbar  \to \rm Z\bbbar$,
$\rm gg  \to \rm Z\bbbar$;
2) $2\to 2$ electroweak subprocesses:
$\qqbar^{'}\to \rm W^\pm Z $,
$\qqbar\to (\rm Z/\gamma^{*}) (\rm Z/\gamma^{*})$;
3) single top quark production
$ \qqbar^{'}\to \rm t\bar{b} \ \ \mbox{or} \ \  \rm \bar{t}b$.
Single top background  has not  been taken  into account in the
previous studies. This contribution becomes more important with growth
of the invariant  \bbbar-pair mass.

Signal and background kinematical  distributions differ mainly due to
concentration of events of QCD backgrounds  mostly in the region of
small $p_\bot$.  We study b quark and charged lepton distributions over
transverse momentum  $p_\bot(\rm b)$ and $p_\bot(\ell)$ as well as
{\bbbar} distributions over transverse momentum $p_\bot(\bbbar)$ and
missing energy $\met$.  It turned out that pair characteristic 
$p_\bot(\bbbar)$ related to high transverse momenta of Higgs boson is
very important for background suppression. After detailed comparison of
the signal and the background distributions  appropriate set of
cuts has been found:
1) $ p_\bot(\rm b), p_\bot(\bbbar)> 20\gev; p_\bot(\ell)> 15\gev$;
2) typical cuts for TEVATRON detectors:
$\met > 20\gev$; $|\eta_{\rm b}| < 1.5$; $|\eta_\ell| < 2$;
$|\Delta R_{\rm \bar{b}}| > 0.7$; $|\Delta R_{\rm b\ell}| > 0.7$;
3)  $-2\Delta M_{\rm \bar{b}}+M_{\rm \bar{b}} < M_{\rm \bar{b}} < 2\Delta
M_{\rm \bar{b}}+ M_{\rm \bar{b}}$.
 This set  of cuts reduces the total background more than 40 times.
{\small
\begin{table}
\hspace*{-0.6cm} 
\begin{tabular}{c c c l l }
\multicolumn{2}{c}{$\sqrt{s}=2$ TeV}&&&\\
 \hline
 $M_{H}$ &$W^\pm H$ &$ZH$ &$Wb\bar{b}+WZ+b\bar{t}$ &$Zb\bar{b}+ZZ$ \\ 
\hline
 60           & 35  & 27  & 53\ \ +\ 1 \ \ +\ 8& 41\ \ +\ 0   \\
 80           & 27  & 19  & 45\ \ +\ 43\ \ +\ 12& 40\ \ +\ 28  \\
 100          & 17  & 12  & 35\ \ +\ 44\ \ +\ 12& 32\ \ +\ 29  \\
 120          & 9   & 7   & 26\ \ +\ 1 \ \ +\ 13& 25\ \ +\ 1   \\
 140          & 4   & 4   & 18\ \ +\ 0 \ \ +\ 11& 17\ \ +\ 0   \\ 
\multicolumn{2}{c}{$\sqrt{s}=4$ TeV}&&&\\
\hline
 60           & 57  & 40  & 82\ \ +\ 3 \ \ +\  18& 122\ \ +\ 1   \\
 80           & 43  & 30  & 74\ \ +\ 83\ \ +\ 15& 138\ \ +\ 69  \\
 100          & 29  & 21  & 62\ \ +\ 83\ \ +\ 18& 112\ \ +\ 71  \\
 120          & 18  & 13  & 46\ \ +\ 3 \ \ +\ 21& 88 \ \ +\ 4   \\
 140          & 8   & 7   & 37\ \ +\ 0 \ \ +\ 20& 69 \ \ +\ 1   \\ 
\hline
\end{tabular}
\vspace*{-0.0cm} 
\caption{\label{higgs-t}
Number of events for Higgs boson signal and background after
cuts}
\vspace*{-0.3cm} 
\end{table}}
The number of events corresponding to  signal and background after the
cuts application is presented in  Table~\ref{higgs-t}. Requiring 
3-standard deviation criteria one can find Higgs up to the mass $\simeq
100\gev$ at $\sqs=2\tev$ and $\simeq 120\gev$ at $\sqs=4\tev$.  These
numbers are obtained under assumption of $50\%$ efficiency of double 
b-tagging. 
\begin{figure}[t]
\vspace*{-1.0cm} 
\begin{center} 
\hspace*{-0.5cm} 
\epsfxsize=10cm  
\epsffile{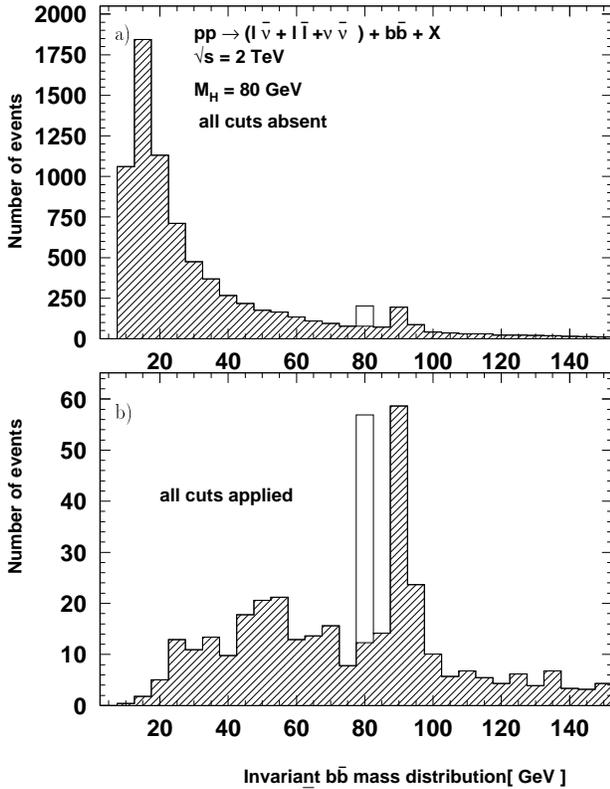} 
\vspace*{-2.5cm} 
\end{center}
\vspace*{-0.3cm} 
\caption{\label{higgs-bb} Invariant $b\bar{b}$ mass distribution before and
after
cuts} 
\vspace*{-0.4cm} 
\end{figure} 
One can see  that with the decreasing  Higgs boson mass the
contribution of the single top processes  to the total background  becomes 
more significant. It should be noted also that $\rm Wbb$ background
increases approximately  two times with  the increasing  of $\sqs$
from  2 to $4\tev$ as well as $\rm W^\pm H$ (or $\rm ZH$) signal.  At the same time 
$\rm Zbb$ cross section becomes about  3-4 times higher because the   
$\rm gg\to Zbb$ subprocess contribution grows  about 6  times.  That is 
why  increasing of the collider energy from 2 to $4\tev$ extends  Higgs
mass interval only up to  $120\gev$.

In Fig.~\ref{higgs-bb} we demonstrate the effective \bbbar-invariant
mass distribution for the cases $M_{\rm H}=80\gev$, $\sqs=2\tev$
without any cuts and with cuts applied revealing clear peak from Higgs
boson. 
\vspace*{-0.2cm} 
\section*{Conclusions}
\vspace*{-0.2cm}
Study of single top quark and Higgs boson
production  at the upgraded Tevatron is  
not only important but also convenient since they  have very similar signatures.

It was shown that the signal from the single top quark can be
extracted  and  therefore  {\wtb} vertex and {\vtb} CKM parameter at
the upgraded TEVATRON can be measured. The main backgrounds have been
studied: {\wbbbar}, {\wjj}, ${\jbbbar}+{\jjbbbar}$ and  the set of
kinematical cuts on $p_\bot(\rm jet)$, $\sqrt{\hat{s}}$, $p_\bot(\rm
W)$, di-jet mass and $H_\bot$ variables has been worked out leading to
the signal-to-background ratio $\simeq 1/3$.

Analysis of the 4-fermion final state reactions for Higgs boson signal
and it's background ($\rm WZ$, $\rm ZZ$,  gluonic background 
{\wbbbar}, $\rm Z\bbbar$ and  single top quark) shows the possibility to
find the last unobserved SM particle up to the mass  $100\gev$   at
$\sqs = 2\tev$ and up to about $120\gev$  at $\sqs = 4\tev$.

{\bf Acknowledgments}.
One of us (A.B.) is grateful to the Organizing Committee for warm hospitality and 
financial support.The financial support of the Russian Foundation for Basic
Research   is acknowledged.

\end{document}